\documentclass[graybox]{svmult}

\usepackage{mathptmx}       % selects Times Roman as basic font
\usepackage{helvet}         % selects Helvetica as sans-serif font
\usepackage{courier}        % selects Courier as typewriter font
\usepackage{type1cm}        % activate if the above 3 fonts are
\usepackage{makeidx}         % allows index generation
\usepackage{graphicx}        % standard LaTeX graphics tool
\usepackage{multicol}        % used for the two-column index
\usepackage[bottom]{footmisc}% places footnotes at page bottom

\usepackage{amsmath,amssymb}

\makeindex             % used for the subject index
\begin{document}

\title*{Feynman Graphs}
\author{Stefan Weinzierl}
\institute{Stefan Weinzierl \at PRISMA Cluster of Excellence, Institut f{\"u}r Physik, Johannes Gutenberg-Universit{\"a}t Mainz, D - 55099 Mainz, Germany, \email{stefanw@thep.physik.uni-mainz.de}
}
\maketitle

\abstract{
In these lectures I discuss Feynman graphs and the associated Feynman integrals.
Of particular interest are the classes functions, which appear in the evaluation of Feynman integrals.
The most prominent class of functions is given by multiple polylogarithms.
The algebraic properties of multiple polylogarithms are reviewed in the second part of these lectures.
The final part of these lectures is devoted to Feynman integrals, which cannot be expressed in terms of multiple polylogarithms.
Methods from algebraic geometry provide tools to tackle these integrals.
}

% -----------------------------------------------------------------------------------
\section{Feynman graph polynomials}
\label{weinzierl_section_graph_polynomials}

\index{graph polynomial}
The first part of these lectures is centred around two graph polynomials.
We will give four different definitions of these two polynomials, each definition will shed a different light on the nature of these
polynomials.
The presentation in this section follows \cite{Bogner:2010kv}.

\subsection{Graphs}
\label{weinzierl_section_graphs}

Let us start with a few basic definitions:
A graph consists of edges and vertices.
We will mainly consider connected graphs.
The valency of a vertex is the number of edges attached to it.
Vertices of valency $0$, $1$ and $2$ are special.
A vertex of valency $0$ is necessarily disconnected from the rest of graph and therefore not relevant for connected graphs.
A vertex of valency $1$ has exactly one edge attached to it. This edge is called an external edge. All other edges are called internal edges.
In the physics community it is common practice not to draw a vertex of valency 1, but just the external edge.
A vertex of valency $2$ is called a mass insertion and is usually not considered.
Therefore in physics it is usually implied that a genuine vertex has a valency of three or greater.

An edge in a Feynman graph represents a propagating particle. The edges are drawn in a way as to represent the different types of particles.
For example, one uses lines with an arrow for fermions, wavy lines for photons or curly lines for gluons. 
A simple line without decorations is used for scalar particles.
To each (orientated) edge we associate a $D$-dimensional vector $q$ and a number $m$, describing the momentum and the mass of the particle. $D$ is the dimensions of space-time.

Vertices of valency $n\ge 3$ represent interactions of $n$ particles.
At each vertex we have momentum conservation: The sum of all momenta flowing into the vertex equals the sum of all momenta flowing out
of the vertex.

To each Feynman graph we can associate a new graph, obtained by replacing each propagator of the original graph by a scalar propagator.
This new graph is called the underlying topology. This new graph does no longer carry the information on the type of the particles propagating
along the edges.
We will later associate to each Feynman graph an integral, called the Feynman integral of this graph.
It turns out, that the Feynman integral corresponding to an arbitrary Feynman graph can always be expressed as a linear combination of 
Feynman integrals corresponding to Feynman graphs with scalar propagators.
Therefore it is sufficient to restrict ourselves to the underlying topology and to restrict our study to Feynman graphs with scalar propagators.

Let us now consider a graph $G$ with $n$ edges and $r$ vertices. Assume that the graph has $k$ connected components.
The loop number $l$ is defined by
\begin{eqnarray}
 l & = & n-r+k.
\end{eqnarray}
If the graph is connected we have $l=n-r+1$.
The loop number $l$ is also called the first Betti number of the graph or the cyclomatic number.
In the physics context it has the following interpretation:
If we fix all momenta of the external lines and if we impose momentum conservation at each vertex, then the loop number is equal to the number of
independent momentum vectors not constrained by momentum conservation.

A connected graph of loop number $0$ is called a tree.
A graph of loop number $0$, connected or not, is called a forest.
If the forest hast $k$ connected components, it is called a $k$-forest.
A tree is a $1$-forest.

\subsection{Spanning forests}
\label{weinzierl_section_spanning_forests}

\index{spanning forest}
Given an arbitrary connected graph $G$, a spanning tree of $G$ is a subgraph, which contains all the vertices of $G$ and which is a tree.
In a similar way,
given an arbitrary connected graph $G$, a spanning $k$-forest of $G$ is a subgraph, which contains all the vertices of $G$ and which is a $k$-forest.

We have already associated to each edge a momentum vector and a mass. 
In addition we associate now to each internal edge $e_j$ a real (or complex) variable $x_j$.
The variables $x_j$ are called Feynman parameters.
For each graph we can define two polynomials ${\mathcal U}$ and ${\mathcal F}$
in the variables $x_j$ as follows:
Let $G$ be a connected graph and ${\cal T}_1$ the set of its spanning trees.
The first graph polynomial is defined by
\begin{eqnarray}
 {\mathcal U} & = &
 \sum\limits_{T \in {\mathcal T}_1} \prod\limits_{e_j \notin T} x_j.
\end{eqnarray}
This is best illustrated with an example.
\begin{figure}[t]
\sidecaption[t]
\includegraphics[bb= 125 610 270 700, scale=.65]{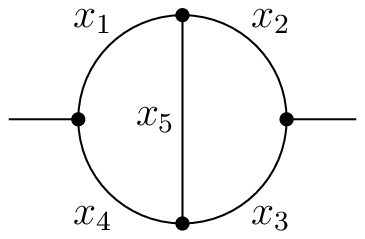}
\includegraphics[bb= 140 570 475 685, scale=.65]{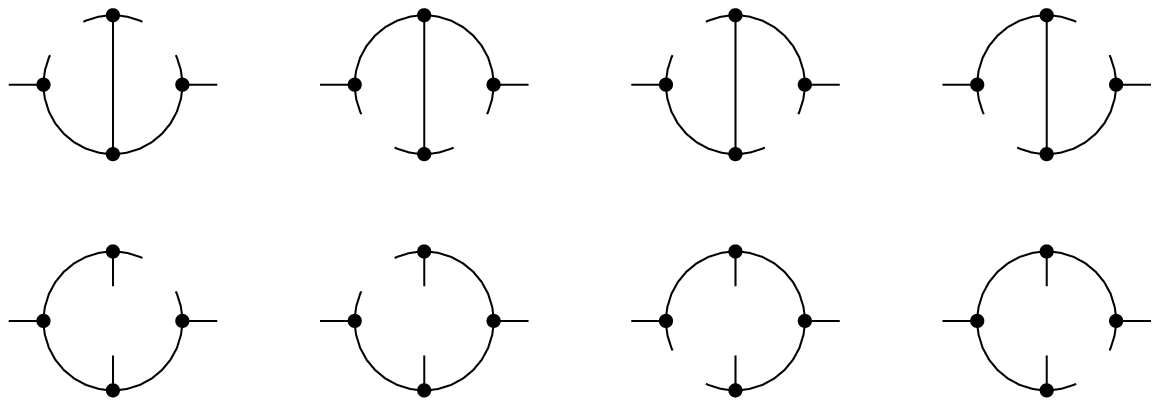}
\caption{An example of a Feynman graph and the associated set of spanning trees.}
\label{weinzierl_fig1}
\end{figure}
Fig.~(\ref{weinzierl_fig1}) shows a Feynman graph decorated with the Feynman parameters $x_1$ to $x_5$, as well as
the associated set ${\mathcal T}_1$ of spanning trees.
For each spanning tree we take the Feynman parameters associated to the edges not belonging to this spanning tree.
Summing over all spanning trees we obtain for this example
\begin{eqnarray}
 {\mathcal U} & = &
  x_1 x_2
 + x_3 x_4
 + x_1 x_3
 + x_2 x_4
 + x_2 x_5
 + x_1 x_5
 + x_4 x_5
 + x_3 x_5.
\end{eqnarray}
${\mathcal U}$ is also called the first Symanzik polynomial of the graph $G$.
In mathematics, the Kirchhoff polynomial of a graph is better known.
It is defined by
\begin{eqnarray}
 {\mathcal K}
 & = & 
 \sum\limits_{T\in {\mathcal T}_1} 
     \prod\limits_{e_j \in T} x_j.
\end{eqnarray}
The difference between the two definitions is given by the fact that in the case of ${\mathcal K}$ we consider all
edges belonging to the spanning tree $T$, while in the case of ${\mathcal U}$ we consider all edges not belonging to $T$.
There is a simple relation between the Kirchhoff polynomial ${\mathcal K}$ and 
the first Symanzik polynomial ${\mathcal U}$:
\begin{eqnarray}
{\mathcal U}(x_1,...,x_n)
& = & 
x_1 ... x_n {\mathcal K}\left(\frac{1}{x_1},...,\frac{1}{x_n}\right).
\end{eqnarray}
We now turn to the definition of ${\mathcal F}$.
Let $G$ be a connected graph and ${\mathcal T}_2$ the set of its spanning $2$-forests.
An element of ${\mathcal T}_2$ is denoted as $(T_1,T_2)$.
Let us further denote by $P_{T_i}$ the set of external momenta of $G$ attached to $T_i$.
We first define a polynomial ${\mathcal F}_0$ by
\begin{eqnarray}
\label{weinzierl_def_F0}
 {\mathcal F}_0
 & = & 
 \sum\limits_{(T_1,T_2)\in {\mathcal T}_2} \;
     \left( \prod\limits_{e_i\notin (T_1,T_2)} x_i \right) 
     \left( \sum\limits_{p_j\in P_{T_1}} \sum\limits_{p_k\in P_{T_2}} \frac{p_j \cdot p_k}{\mu^2} \right).
\end{eqnarray}
Here, $p_j \cdot p_k$ is the Minkowski scalar product of two momenta vectors.
$\mu$ is an arbitrary scale introduced to make the expression dimensionless.
${\mathcal F}$ is defined by
\begin{eqnarray}
 {\mathcal F} & = & 
 {\mathcal F}_0 + {\mathcal U} \sum\limits_{i=1}^n x_i \frac{m_i^2}{\mu^2}.
\end{eqnarray}
$m_i$ denotes the mass of the $i$-th internal line.
If all internal masses are zero, we have ${\mathcal F}={\mathcal F}_0$.
${\mathcal F}$ is called the second Symanzik polynomial.
\begin{figure}[t]
\sidecaption[t]
\includegraphics[bb= 110 595 290 730, scale=.65]{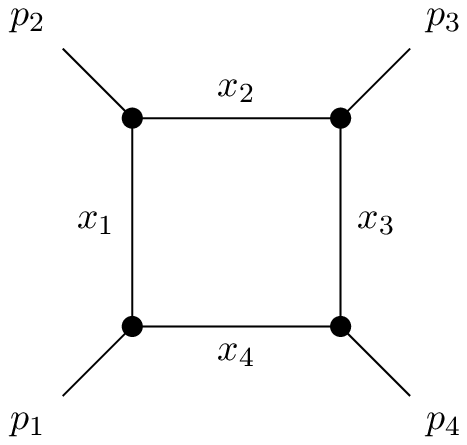}
\includegraphics[bb= 160 543 450 695, scale=.65]{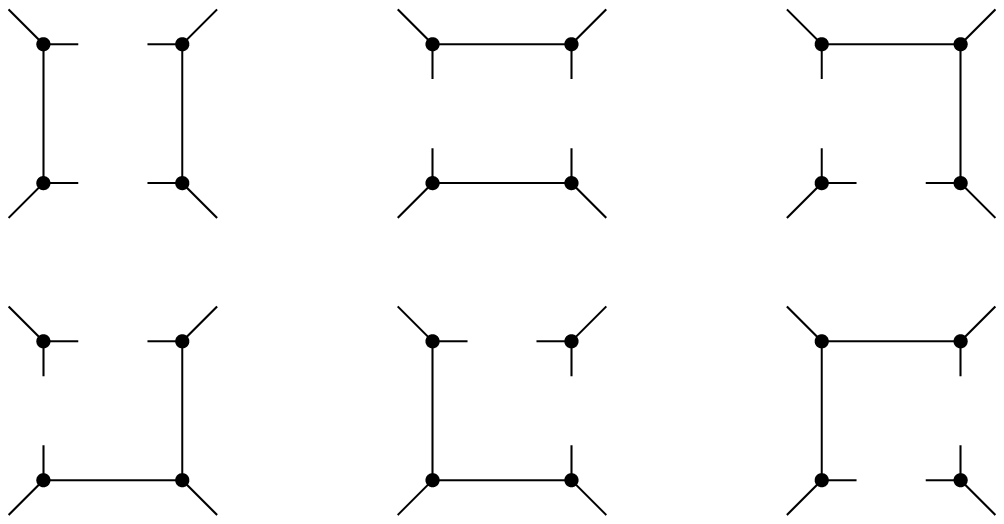}
\caption{An example of a Feynman graph and the associated set of spanning $2$-forests.}
\label{weinzierl_fig2}
\end{figure}
Again, let us illustrate the definition of ${\mathcal F}$ with an example.
Fig.~(\ref{weinzierl_fig2}) shows a Feynman graph and the associated set ${\mathcal T}_2$ of $2$-forests.
For simplicity we assume that all internal masses are zero, therefore we have ${\mathcal F}={\mathcal F}_0$.
In fig.~(\ref{weinzierl_fig2}) we have labelled the internal edges with the Feynman parameters $x_1$ to $x_4$.
The external edges have been labelled with the external momenta $p_1$ to $p_4$.
We orientate these edges such that $p_1$ to $p_4$ are all flowing outwards.
With this choice momentum conservation reads
\begin{eqnarray}
 p_1 + p_2 + p_3 + p_4 & = & 0.
\end{eqnarray}
The Mandelstam variables $s$  and $t$ are defined by
\begin{eqnarray}
 s = \left(p_1+p_2\right)^2,
 & &
 t = \left(p_2+p_3\right)^2.
\end{eqnarray}
From the definition in eq.~(\ref{weinzierl_def_F0}) we find for the polynomial ${\mathcal F}$ for this example
\begin{eqnarray}
{\mathcal F} = 
 x_2 x_4 \frac{(-s)}{\mu^2}
 + x_1 x_3 \frac{(-t)}{\mu^2}
 + x_1 x_4 \frac{(-p_1^2)}{\mu^2}
 + x_1 x_2 \frac{(-p_2^2)}{\mu^2}
 + x_2 x_3 \frac{(-p_3^2)}{\mu^2}
 + x_3 x_4 \frac{(-p_4^2)}{\mu^2}.
 \nonumber
\end{eqnarray}
A few remarks on the two Symanzik polynomials are in order:
Both polynomials are homogeneous in the Feynman parameters, ${\mathcal U}$ is of degree $l$, ${\mathcal F}$ is of degree
$l+1$.
The polynomial ${\mathcal U}$ is linear in each Feynman parameter.
If all internal masses are zero, then
also ${\mathcal F}$ is linear in each Feynman parameter.
In expanded form each monomial of ${\mathcal U}$ has coefficient $+1$.

\subsection{Feynman integrals}
\label{weinzierl_section_feynman_integrals}

\index{Feynman integrals}
\index{Feynman diagram}
Feynman graphs have been invented as a pictorial notation for mathematical expressions arising in the context
of perturbative quantum field theory. 
Each part in a Feynman graph corresponds to a specific expression and the full Feynman graph corresponds to the product of these
expressions.
For scalar theories the correspondence is as follows:
An internal edge corresponds to a propagator
\begin{eqnarray}
 \frac{i}{q^2-m^2},
\end{eqnarray}
an external edge to the factor $1$.
A vertex corresponds in scalar theories also to the factor $1$.
In addition, there is for each internal momentum not constrained by momentum conservation an
integration
\begin{eqnarray} 
 \int \frac{d^Dk}{\left(2\pi\right)^D}.
\end{eqnarray}
Let us now consider a Feynman graph $G$ with $m$ external edges, $n$ internal edges and $l$ loops.
To each internal edge we associate apart from its momentum and its mass a positive integer number $\nu$, giving the power to which the propagator
occurs.
(We can think of $\nu$ as the relict of neglecting vertices of valency $2$. A number $\nu>1$ corresponds to $\nu-1$ mass insertions on this edge).
The momenta flowing through the internal lines can be expressed through the independent loop momenta
$k_1$, ..., $k_l$ and the external momenta $p_1$, ..., $p_m$ as 
\begin{eqnarray}
 q_i & = & \sum\limits_{j=1}^l \rho_{ij} k_j + \sum\limits_{j=1}^m \sigma_{ij} p_j,
 \;\;\;\;\;\; 
 \rho_{ij}, \sigma_{ij} \in \{-1,0,1\}.
\end{eqnarray}
We define the Feynman integral by
\begin{eqnarray}
I_G  & = &
 \frac{\prod\limits_{j=1}^{n}\Gamma(\nu_j)}{\Gamma(\nu-lD/2)}
 \left( \mu^2 \right)^{\nu-l D/2}
 \int \prod\limits_{r=1}^{l} \frac{d^Dk_r}{i\pi^{\frac{D}{2}}}\;
 \prod\limits_{j=1}^{n} \frac{1}{(-q_j^2+m_j^2)^{\nu_j}},
\end{eqnarray}
with $\nu=\nu_1+...+\nu_n$.
The prefactor in front of the integral is the convention used in this article.
This choice is motivated by the fact that after Feynman parametrisation we obtain a simple formula.
Feynman parametrisation makes use of the identity
\begin{eqnarray}
\label{weinzierl_feynman_parametrisation}
 \prod\limits_{j=1}^{n} \frac{1}{P_{j}^{\nu_j}} 
 & = &
 \frac{\Gamma\left(\nu\right)}{\prod\limits_{i=1}^n \Gamma\left(\nu_j\right)}
 \int\limits_\Delta \omega
 \left( \prod\limits_{i=1}^n x_i^{\nu_i-1} \right)
 \left( \sum\limits_{j=1}^{n} x_{j} P_{j} \right)^{-\nu},
\end{eqnarray}
where $\omega$ is a differential $(n-1)$-form given by
\begin{eqnarray}
 \omega & = & \sum\limits_{j=1}^n (-1)^{j-1}
  \; x_j \; dx_1 \wedge ... \wedge \widehat{dx_j} \wedge ... \wedge dx_n.
\end{eqnarray}
The hat indicates that the corresponding term is omitted.
The integration is over
\begin{eqnarray}
 \Delta & = & \left\{ \left[ x_1 : x_2 : ... : x_n \right] \in {\mathbb P}^{n-1} | x_i \ge 0, 1 \le i \le n \right\}.
\end{eqnarray}
We use eq.~(\ref{weinzierl_feynman_parametrisation}) with $P_j=-q_j^2+m_j^2$.
We can write
\begin{eqnarray}
 \sum\limits_{j=1}^{n} x_{j} (-q_j^2+m_j^2)
 & = & 
 - \sum\limits_{r=1}^{l} \sum\limits_{s=1}^{l} k_r M_{rs} k_s + \sum\limits_{r=1}^{l} 2 k_r \cdot Q_r - J,
\end{eqnarray}
where $M$ is a $l \times l$ matrix with scalar entries and $Q$ is a $l$-vector
with $D$-vectors as entries.
After Feynman parametrisation the integrals over the loop momenta $k_1$, ..., $k_l$ can be done and we obtain
\begin{eqnarray}
\label{weinzierl_feynman_integral}
I_G  & = &
 \int\limits_{\Delta}  \omega
 \left( \prod\limits_{j=1}^n x_j^{\nu_j-1} \right)
 \frac{{\mathcal U}^{\nu-(l+1) D/2}}{{\mathcal F}^{\nu-l D/2}}.
\end{eqnarray}
The functions ${\mathcal U}$ and ${\mathcal F}$ are given by
\begin{eqnarray}
\label{weinzierl_second_definition_U_and_F}
 {\mathcal U} = \mbox{det}(M),
 & &
 {\mathcal F} = \mbox{det}(M) \left( - J + Q M^{-1} Q \right)/\mu^2.
\end{eqnarray} 
It can be shown that eq.~(\ref{weinzierl_second_definition_U_and_F}) agrees with the definition of ${\mathcal U}$ and ${\mathcal F}$ given in 
section~\ref{weinzierl_section_spanning_forests} in terms of spanning trees and spanning forests.
Thus, eq.~(\ref{weinzierl_second_definition_U_and_F}) provides a second definition of the two graph polynomials.
Eq.~(\ref{weinzierl_feynman_integral}) defines the Feynman integral of a graph $G$ in terms of the two graph polynomials ${\mathcal U}$ and ${\mathcal F}$.
A few remarks are in order:
The integral over the Feynman parameters is a $(n-1)$-dimensional integral in projective space ${\mathbb P}^{n-1}$, where
$n$ is the number of internal edges of the graph.
Singularities may arise if the zero sets of ${\cal U}$ and ${\cal F}$ intersect the region of integration.
The dimension $D$ of space-time enters only in the exponents of the integrand and 
the exponents act as a regularisation.

\subsection{The Laplacian of a graph}
\label{weinzierl_section_laplacian}

\index{Laplacian}
For a graph $G$ with $n$ edges and $r$ vertices define the Laplacian $L$ \cite{Tutte:1984,Stanley:1998} as a
$r \times r$-matrix with 
\begin{eqnarray}
 L_{ij} & = & \left\{ \begin{array}{ll} 
                      \sum x_k & \mbox{if $i =j$ and edge $e_k$ is attached to $v_i$ and is not a self-loop,} \\
                      - \sum x_k & \mbox{if $i \neq j$ and edge $e_k$ connects $v_i$ and $v_j$.}\\
                      \end{array} \right.
 \nonumber
\end{eqnarray}
We speak of a self-loop (or tadpole) if an edge starts and ends at the same vertex.
In the sequel we will need minors of the matrix $L$ and it is convenient to introduce the
following notation:
For a $r \times r$ matrix $A$ we denote by $A[i_1,...,i_k;j_1,...,j_k]$ the
$(r-k)\times(r-k)$ matrix, which is obtained from $A$ by deleting the rows $i_1$, ..., $i_k$ and the 
columns $j_1$, ..., $j_k$.
For $A[i_1,...,i_k;i_1,...,i_k]$ we will simply write $A[i_1,...,i_k]$.
The matrix-tree theorem relates the Laplacian of a graph to its Kirchhoff polynomial:
\begin{eqnarray}
 {\mathcal K} & = & \det L[i].
\end{eqnarray}
A generalisation by the all-minor matrix tree theorem \cite{Chaiken:1982,Chen:1982,Moon:1994}
leads to the following expressions for the graph polynomials ${\mathcal U}$
and ${\mathcal F}_0$:
Starting from a graph $G$ with $n$ internal edges, $r$ internal vertices $(v_1,...,v_r)$ and $m$ external legs, we first attach
$m$ additional vertices $(v_{r+1},...,v_{r+m})$ to the ends of the external legs and then 
associate the parameters $z_1$, ..., $z_m$ with the external edges.
This defines a new graph $\tilde{G}$.
We now consider the Laplacian $\tilde{L}$ of $\tilde{G}$ and the
polynomial
\begin{eqnarray}
 {\mathcal W}(x_1,...,x_n,z_1,...,z_m) & = & \det \; \tilde{L}[r+1,...,r+m].
\end{eqnarray}
We then expand ${\mathcal W}$ in polynomials homogeneous in the variables $z_j$:
\begin{eqnarray}
 {\mathcal W} & = & {\mathcal W}^{(0)} + {\mathcal W}^{(1)} + {\mathcal W}^{(2)} + ... + {\mathcal W}^{(m)},
 \nonumber \\
 {\mathcal W}^{(k)} & = &
 \sum\limits_{1\le j_1<...<j_k \le m} {\mathcal W}^{(k)}_{(j_1,...,j_k)}(x_1,...,x_n) \; z_{j_1} ... z_{j_k}.
\end{eqnarray}
We then have
\begin{eqnarray}
 0 & = & {\mathcal W}^{(0)},
 \nonumber \\
 {\mathcal U} & = & x_1 ... x_n \; {\mathcal W}^{(1)}_{(j)}\left(\frac{1}{x_1},...,\frac{1}{x_n}\right)
 \;\;\;\;\;\;\;\;\; \mbox{for any $j$},
 \nonumber \\
 {\mathcal F}_0 & = & 
   x_1 ... x_n \sum\limits_{(j,k)} 
   \left( \frac{p_j \cdot p_k}{\mu^2} \right)
   \cdot 
   {\mathcal W}^{(2)}_{(j,k)}\left(\frac{1}{x_1},...,\frac{1}{x_n}\right).
\end{eqnarray}
This provides a third definition of the Feynman graph polynomials ${\cal U}$ and ${\cal F}$.
This formulation is particularly well suited for computer algebra.

\subsection{Deletion and contraction properties}
\label{weinzierl_section_deletion_contraction}

\begin{figure}[t]
\sidecaption[t]
\includegraphics[bb= 170 645 440 710, scale=.65]{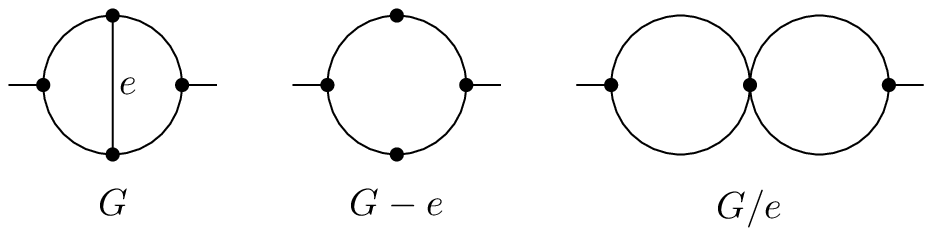}
\caption{A graph $G$, together with the graph $G-e$, where the edge $e$ has been deleted and the graph $G/e$, where the edge $e$ has been contracted.}
\label{weinzierl_fig4}
\end{figure}

\index{deletion}
\index{contraction}
Let us now consider a recursive definition of the two graph polynomials based on deletion and contraction properties.
We first define a regular edge to be an edge, which is neither a self-loop nor a bridge.
In graph theory an edge is called a bridge, if the deletion of the edge increases
the number of connected components. 
\begin{figure}[t]
\sidecaption[t]
\includegraphics[bb= 187 650 425 720, scale=.65]{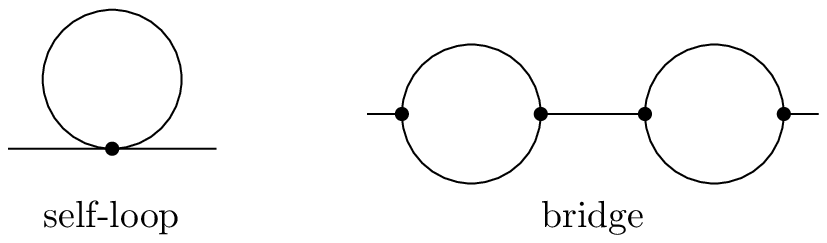}
\caption{Examples of graphs containing a self-loop (left) or a bridge (right).}
\label{weinzierl_fig3}
\end{figure}
Examples for graphs containing either a self-loop or a bridge are shown in fig.~(\ref{weinzierl_fig3}).
For a graph $G$ and a regular edge $e$ we define
\begin{eqnarray}
G/e & & \mbox{to be the graph obtained from $G$ by contracting the regular edge $e$,} \nonumber \\
G- e & & \mbox{to be the graph obtained from $G$ by deleting the regular edge $e$.}
\end{eqnarray}
The operations of deletion and contraction are illustrated in fig.~(\ref{weinzierl_fig4}).
For any regular edge $e_k$ we have
\begin{eqnarray}
\label{weinzierl_recursion_U_F0}
 {\mathcal U}(G) & = & {\mathcal U}(G/e_k) + x_k {\mathcal U}(G-e_k), 
 \nonumber \\
 {\mathcal F}_0(G) & = & {\mathcal F}_0(G/e_k) + x_k {\mathcal F}_0(G-e_k).
\end{eqnarray}
The recursion terminates when all edges are either bridges or self-loops.
These graphs are called terminal forms.
If a terminal form has $r$ vertices and $l$ (self-) loops, 
then there are $(r-1)$ ``tree-like'' propagators, where the momenta flowing through these propagators
are linear combinations of the external momenta $p_i$ alone and independent of the independent loop momenta $k_j$.
The momenta of the remaining $l$ propagators are on the other hand independent 
of the external momenta and can be taken as the independent loop momenta $k_j$, $j=1,...,l$.
Let us agree that we label 
the $(r-1)$ ``tree-like'' edges from $1$ to $r-1$, and the remaining $l$
edges by $r$, ..., $n$ with $n=r+l-1$.
We further denote the momentum squared flowing through edge $j$ by $q_j^2$.
For a terminal form we have
\begin{eqnarray}
\label{weinzierl_terminus_U_F0}
 {\mathcal U} = x_{r} ... x_{n},
 & &
 {\mathcal F}_0 = x_{r} ... x_{n} \sum\limits_{j=1}^{r-1} x_j \left( \frac{-q_j^2}{\mu^2} \right).
\end{eqnarray}
Eq.~(\ref{weinzierl_recursion_U_F0}) together with eq.~(\ref{weinzierl_terminus_U_F0})
provides a fourth definition of the graph polynomials ${\mathcal U}$ and ${\mathcal F}$.

Let $e_a$ and $e_b$ be two regular edges, which share a common vertex.
We have the following factorisation theorems:
\begin{eqnarray}
\label{weinzierl_factorisation_U_U_and_U_F0}
\lefteqn{
 {\mathcal U}\left( G/e_a- e_b \right) {\mathcal U}\left( G/e_b- e_a \right)
 -
 {\mathcal U}\left( G- e_a-e_b \right) {\mathcal U}\left( G/e_a/e_b \right)
 = 
 \left( \frac{\Delta_1}{x_a x_b} \right)^2,
} & &
 \nonumber \\
 \lefteqn{
 {\mathcal U}\left( G/e_a- e_b \right) {\mathcal F}_0\left( G/e_b- e_a \right)
 -
 {\mathcal U}\left( G- e_a-e_b \right) {\mathcal F}_0\left( G/e_a/e_b \right)
} 
&& \nonumber \\
\lefteqn{
 +
 {\mathcal F}_0\left( G/e_a- e_b \right) {\mathcal U}\left( G/e_b- e_a \right)
 -
 {\mathcal F}_0\left( G- e_a-e_b \right) {\mathcal U}\left( G/e_a/e_b \right)
 = 
} & & \nonumber \\
 & & 
 2 \left( \frac{\Delta_1}{x_a x_b} \right) \left( \frac{\Delta_2}{x_a x_b} \right).
 \hspace*{70mm}
\end{eqnarray}
$\Delta_1$ and $\Delta_2$ are polynomials in the Feynman parameters and can be expressed as sums over $2$-forests 
and sums over $3$-forests, respectively \cite{Bogner:2010kv}.
If for all external momenta one has
\begin{eqnarray}
 \left( p_{i_1} \cdot p_{i_2} \right) \cdot \left( p_{i_3} \cdot p_{i_4} \right)
 & = & 
 \left( p_{i_1} \cdot p_{i_3} \right) \cdot \left( p_{i_2} \cdot p_{i_4} \right),
\end{eqnarray}
then
\begin{eqnarray}
 {\mathcal F}_0\left( G/e_a- e_b \right) {\mathcal F}_0\left( G/e_b- e_a \right)
 -
 {\mathcal F}_0\left( G- e_a-e_b \right) {\mathcal F}_0\left( G/e_a/e_b \right)
 = 
 \left( \frac{\Delta_2}{x_a x_b} \right)^2.
 \nonumber
\end{eqnarray}
The factorisation theorems follow from Dodgson's identity \cite{Dodgson:1866,Zeilberger:1997}, 
which states that for any $n\times n$ matrix $A$ one has
\begin{eqnarray}
 \det\left( A \right) \det\left( A[i,j] \right)
 & = & 
 \det\left( A[i] \right) \det\left( A[j] \right)
 -
 \det\left( A[i;j] \right) \det\left( A[j;i] \right).
\end{eqnarray}
We recall that
\begin{eqnarray}
 A[i] & & \mbox{is obtained from $A$ by deleting the $i$-th row and column,}
 \nonumber \\
 A[i;j] & & \mbox{is obtained from $A$ by deleting the $i$-th row and the $j$-th column,}
 \nonumber \\
 A[i,j] & & \mbox{is obtained from $A$ by deleting the rows and columns $i$ and $j$.}
 \nonumber
\end{eqnarray}
The first formula of eq.~(\ref{weinzierl_factorisation_U_U_and_U_F0}) is at the heart of the reduction algorithm of
\cite{Brown:2008,Brown:2009a}.

% -----------------------------------------------------------------------------------
\section{Multiple polylogarithms}
\label{weinzierl_section_multiple_polylogs}

\index{series expansion}
Let us come back to the Feynman integrals defined in eq.~(\ref{weinzierl_feynman_integral}).
A Feynman integral has an expansion as a Laurent series in the parameter $\varepsilon=(4-D)/2$ of dimensional regularisation:
\begin{eqnarray}
 I_G & = &
 \sum\limits_{j=-2l}^\infty c_j \varepsilon^j.
\end{eqnarray}
The Laurent series of an $l$-loop integral can have poles in $\varepsilon$ up to the order $(2l)$. The poles in $\varepsilon$ correspond to
ultraviolet or infrared divergences.
The coefficients $c_j$ are functions of the scalar products $p_j \cdot p_k$, the masses $m_i$ and (in a trivial way) of the arbitrary scale $\mu$.
An interesting question is, which functions do occur in the coefficients $c_j$.

\subsection{One-loop integrals}
\label{weinzierl_one_loop}

\index{one-loop integrals}
The question, which functions occur in the coefficients $c_j$ has a satisfactory answer for one-loop integrals.
If we restrict our attention to the coefficients $c_j$ with $j\le 0$ (i.e. to $c_{-2}$, $c_{-1}$ and $c_0$), then these coefficients can be expressed
as a sum of algebraic functions of the scalar products and the masses times two transcendental functions,
whose arguments are again algebraic functions of the scalar products and the masses.

The two transcendental functions are the logarithm and the dilogarithm:
\begin{eqnarray}
\label{weinzierl_log_dilog}
 \mbox{Li}_1(x) & = & \sum\limits_{n=1}^\infty \frac{x^n}{n} = - \ln(1-x),
 \nonumber \\
 \mbox{Li}_2(x) & = & \sum\limits_{n=1}^\infty \frac{x^n}{n^2}.
\end{eqnarray}

\subsection{The sum representation of multiple polylogarithms}
\label{weinzierl_sum_repr_polylogs}

\index{(infinite) nested sum}
\index{symbolic summation}
Beyond one-loop an answer to the above question is not yet known.
We know however that the following generalisations occur:
From eq.~(\ref{weinzierl_log_dilog}) it is not too hard to imagine that the generalisation includes the classical polylogarithms
defined by
\begin{eqnarray}
 \mbox{Li}_m(x) & = & \sum\limits_{n=1}^\infty \frac{x^n}{n^m}.
\end{eqnarray}
However, explicit calculations at two-loops and beyond show that a wider generalisation towards functions of several variables is needed
and one arrives at the 
multiple polylogarithms defined by \cite{Goncharov_no_note,Goncharov:2001,Borwein}
\begin{eqnarray} 
\label{weinzierl_def_multiple_polylogs_sum}
 \mbox{Li}_{m_1,...,m_k}(x_1,...,x_k)
  & = & \sum\limits_{n_1>n_2>\ldots>n_k>0}^\infty
     \frac{x_1^{n_1}}{{n_1}^{m_1}}\ldots \frac{x_k^{n_k}}{{n_k}^{m_k}}.
\end{eqnarray}
Methods for the numerical evaluation of multiple polylogarithms can be found in \cite{Vollinga:2004sn}. 
The values of the multiple polylogarithms at $x_1=...x_k=1$ are called multiple $\zeta$-values \cite{Borwein,Blumlein:2009}:
\index{multiple zeta values}
\begin{eqnarray}
\zeta_{m_1,...,m_k} & = & \mbox{Li}_{m_1,m_2,...,m_k}(1,1,...,1) 
 =
 \sum\limits_{n_1 > n_2 > ... > n_k > 0}^\infty 
 \;\;\;
 \frac{1}{n_1^{m_1}} \cdot ... \cdot \frac{1}{n_k^{m_k}}.
\end{eqnarray}
Important specialisations of multiple polylogarithms 
are the harmonic polylogarithms \cite{Remiddi:1999ew,Gehrmann:2000zt}
\begin{eqnarray}
H_{m_1,...,m_k}(x) & = & \mbox{Li}_{m_1,...,m_k}(x,\underbrace{1,...,1}_{k-1}),
\end{eqnarray}
Further specialisations leads to Nielsen's generalised polylogarithms \cite{Nielsen}
\begin{eqnarray}
S_{n,p}(x) & = & \mbox{Li}_{n+1,1,...,1}(x,\underbrace{1,...,1}_{p-1}).
\end{eqnarray}

\subsection{The integral representation of multiple polylogarithms}
\label{weinzierl_integral_repr_polylogs}

\index{symbolic integration}
In eq.~(\ref{weinzierl_def_multiple_polylogs_sum}) we have defined multiple polylogarithms through the sum representation.
In addition, multiple polylogarithms have an integral representation. 
To discuss the integral representation it is convenient to introduce for $z_k \neq 0$
the following functions
\begin{eqnarray}
\label{weinzierl_Gfuncdef}
G(z_1,...,z_k;y) & = &
 \int\limits_0^y \frac{dt_1}{t_1-z_1}
 \int\limits_0^{t_1} \frac{dt_2}{t_2-z_2} ...
 \int\limits_0^{t_{k-1}} \frac{dt_k}{t_k-z_k}.
\end{eqnarray}
In this definition one variable is redundant due to the following scaling relation:
\begin{eqnarray}
\label{weinzierl_G_scaling_relation}
G(z_1,...,z_k;y) & = & G(x z_1, ..., x z_k; x y)
\end{eqnarray}
If one further defines $g(z;y) = 1/(y-z)$, then one has
\begin{eqnarray}
\frac{d}{dy} G(z_1,...,z_k;y) & = & g(z_1;y) G(z_2,...,z_k;y)
\end{eqnarray}
and
\begin{eqnarray}
\label{weinzierl_Grecursive}
G(z_1,z_2,...,z_k;y) & = & \int\limits_0^y dt \; g(z_1;t) G(z_2,...,z_k;t).
\end{eqnarray}
One can slightly enlarge the set and define $G(0,...,0;y)$ with $k$ zeros for $z_1$ to $z_k$ to be
\begin{eqnarray}
\label{weinzierl_trailingzeros}
G(0,...,0;y) & = & \frac{1}{k!} \left( \ln y \right)^k.
\end{eqnarray}
This permits us to allow trailing zeros in the sequence
$(z_1,...,z_k)$ by defining the function $G$ with trailing zeros via eq.~(\ref{weinzierl_Grecursive}) 
and eq.~(\ref{weinzierl_trailingzeros}).
To relate the multiple polylogarithms to the functions $G$ it is convenient to introduce
the following short-hand notation:
\begin{eqnarray}
\label{weinzierl_Gshorthand}
G_{m_1,...,m_k}(z_1,...,z_k;y)
 & = &
 G(\underbrace{0,...,0}_{m_1-1},z_1,...,z_{k-1},\underbrace{0...,0}_{m_k-1},z_k;y)
\end{eqnarray}
Here, all $z_j$ for $j=1,...,k$ are assumed to be non-zero.
One then finds
\begin{eqnarray}
\label{weinzierl_Gintrepdef}
\mbox{Li}_{m_1,...,m_k}(x_1,...,x_k)
& = & (-1)^k 
 G_{m_1,...,m_k}\left( \frac{1}{x_1}, \frac{1}{x_1 x_2}, ..., \frac{1}{x_1...x_k};1 \right).
\end{eqnarray}
The inverse formula reads
\begin{eqnarray}
G_{m_1,...,m_k}(z_1,...,z_k;y) & = & 
 (-1)^k \; \mbox{Li}_{m_1,...,m_k}\left(\frac{y}{z_1}, \frac{z_1}{z_2}, ..., \frac{z_{k-1}}{z_k}\right).
\end{eqnarray}
Eq.~(\ref{weinzierl_Gintrepdef}) together with eq.~(\ref{weinzierl_Gshorthand}) and eq.~(\ref{weinzierl_Gfuncdef})
defines an integral representation for the multiple polylogarithms.
As an example, we obtain from eq.~(\ref{weinzierl_Gintrepdef}) and eq.~(\ref{weinzierl_G_scaling_relation}) 
the integral representation of harmonic polylogarithms:
\begin{eqnarray}
H_{m_1,...,m_k}(x) & = & 
 \left(-1\right)^k G_{m_1,...,m_k}\left(1,...,1;x\right).
\end{eqnarray}
The function $G_{m_1,...,m_k}(1,...,1;x)$ is an iterated integral in which only the two one-forms
\begin{eqnarray}
 \omega_0 = \frac{dt}{t},
 & &
 \omega_1 = \frac{dt}{t-1}
\end{eqnarray}
corresponding to $z=0$ and $z=1$
appear. If one restricts the possible values of $z$ to zero and the $n$-th roots of unity one arrives at the class of cyclomatic harmonic
polylogarithms \cite{Ablinger:2011te}.

\subsection{Shuffle and quasi-shuffle algebras}
\label{weinzierl_section_shuffle}

\index{shuffle algebra}
Multiple polylogarithms have a rich algebraic structure.  
The representations as iterated integrals and nested sums induce a shuffle algebra and a quasi-shuffle algebra, respectively.
Shuffle and quasi-shuffle algebras are Hopf algebras.
Note that the shuffle algebra of multiple polylogarithms is distinct from the quasi-shuffle algebra of multiple polylogarithms.

Consider a set of letters $A$. 
The set $A$ is called the alphabet.
A word is an ordered sequence of letters:
\begin{eqnarray}
 w & = & l_1 l_2 ... l_k.
\end{eqnarray}
The word of length zero is denoted by $e$.
Let $K$ be a field and consider the vector space of words over $K$.
A shuffle algebra ${\cal A}$ on the vector space of words is defined by
\begin{eqnarray}
\left( l_1 l_2 ... l_k \right) \cdot 
 \left( l_{k+1} ... l_r \right) & = &
 \sum\limits_{\mbox{\tiny shuffles} \; \sigma} l_{\sigma(1)} l_{\sigma(2)} ... l_{\sigma(r)},
\end{eqnarray}
where the sum runs over all permutations $\sigma$, which preserve the relative order of $1,2,...,k$ and of $k+1,...,r$.
The name ``shuffle algebra'' is related to the analogy of shuffling cards: If a deck of cards
is split into two parts and then shuffled, the relative order within the two individual parts
is conserved. A shuffle algebra is also known under the name ``mould symmetral'' \cite{Ecalle}.
The empty word $e$ is the unit in this algebra:
\begin{eqnarray}
 e \cdot w = w \cdot e = w.
\end{eqnarray}
A recursive definition of the shuffle product is given by
\begin{eqnarray}
\label{weinzierl_def_recursive_shuffle}
\left( l_1 l_2 ... l_k \right) \cdot \left( l_{k+1} ... l_r \right) = 
 l_1 \left[ \left( l_2 ... l_k \right) \cdot \left( l_{k+1} ... l_r \right) \right]
+
 l_{k+1} \left[ \left( l_1 l_2 ... l_k \right) \cdot \left( l_{k+2} ... l_r \right) \right].
\end{eqnarray}
It is a well known fact that the shuffle algebra is actually a (non-cocommutative) Hopf algebra \cite{Reutenauer}.
In a Hopf algebra we have in addition to the multiplication and the unit a counit, a comultiplication and a antipode.
The unit in an algebra can be viewed as a map from $K$ to $A$ and multiplication in an algebra
can be viewed as a map from the tensor product $A \otimes A$ to $A$ (e.g. one takes two elements
from $A$, multiplies them and gets one element out). 
The counit is a map from $A$ to $K$, whereas comultiplication is a map from $A$ to
$A \otimes A$.
We will always assume that the comultiplication is coassociative. The general form of the coproduct is
\begin{eqnarray}
\Delta(a) & = & \sum\limits_i a_i^{(1)} \otimes a_i^{(2)},
\end{eqnarray}
where $a_i^{(1)}$ denotes an element of $A$ appearing in the first slot of $A \otimes A$ and
$a_i^{(2)}$ correspondingly denotes an element of $A$ appearing in the second slot.
Sweedler's notation \cite{Sweedler} consists in dropping the dummy index $i$ and the summation symbol:
\begin{eqnarray}
\Delta(a) & = & 
a^{(1)} \otimes a^{(2)}
\end{eqnarray} 
The sum is implicitly understood. This is similar to Einstein's summation convention, except
that the dummy summation index $i$ is also dropped. The superscripts ${}^{(1)}$ and ${}^{(2)}$ 
indicate that a sum is involved.
Using Sweedler's notation, the compatibility between the multiplication and comultiplication is expressed as
\begin{eqnarray}
\label{weinzierl_bialg}
 \Delta\left( a \cdot b \right)
 & = &
\left( a^{(1)} \cdot b^{(1)} \right)
 \otimes \left( a^{(2)} \cdot b^{(2)} \right).
\end{eqnarray}
The antipode $S$ is map from $A$ to $A$, which fulfils
\begin{eqnarray}
a^{(1)} \cdot S\left( a^{(2)} \right)
=
S\left(a^{(1)}\right) \cdot a^{(2)} 
= e \cdot \bar{e}(a).
\end{eqnarray}
With this background at hand we can now state the coproduct, the counit and the antipode for the
shuffle algebra:
The counit $\bar{e}$ is given by:
\begin{eqnarray}
\bar{e}\left( e\right) = 1, \;\;\;
& &
\bar{e}\left( l_1 l_2 ... l_n\right) = 0.
\end{eqnarray}
The coproduct $\Delta$ is given by:
\begin{eqnarray}
\Delta\left( l_1 l_2 ... l_k \right) 
& = & \sum\limits_{j=0}^k \left( l_{j+1} ... l_k \right) \otimes \left( l_1 ... l_j \right).
\end{eqnarray}
The antipode $S$ is given by:
\begin{eqnarray}
S\left( l_1 l_2 ... l_k \right) & = & (-1)^k \; l_k l_{k-1} ... l_2 l_1.
\end{eqnarray}
The shuffle algebra is generated by the Lyndon words.
If one introduces a lexicographic ordering on the letters of the alphabet
$A$, a Lyndon word is defined by the property $w < v$
for any sub-words $u$ and $v$ such that $w= u v$.

\begin{figure}[t]
\sidecaption[t]
\includegraphics[bb= 105 635 480 710, scale=.65]{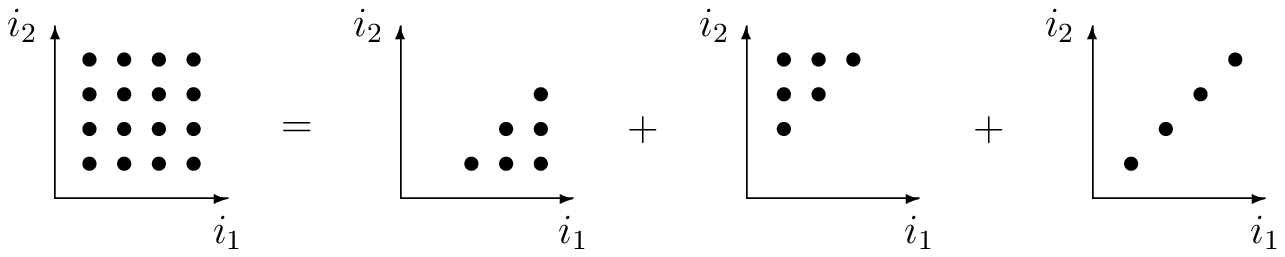}
\caption{A quasi-shuffle algebra follows from replacing the sum over the square by a sum over the lower triangle, a sum over the upper triangle and a sum over the diagonal.}
\label{weinzierl_fig5}
\end{figure}

An important example for a shuffle algebra are iterated integrals.
Let $[a, b]$ be a segment of the real line and $f_1$, $f_2$, ... functions on this interval.
Let us define the following iterated integrals:
\begin{eqnarray}
\label{weinzierl_iterated_integrals}
 I(f_1,f_2,...,f_k;a,b) 
 & = &
 \int\limits_a^b dt_1 f_1(t_1) \int\limits_a^{t_1} dt_2 f_2(t_2) 
 ...
 \int\limits_a^{t_{k-1}} dt_k f_k(t_k) 
\end{eqnarray}
For fixed $a$ and $b$ we have a shuffle algebra:
\begin{eqnarray}
 I(f_1,f_2,...,f_k;a,b) \cdot I(f_{k+1},...,f_r; a,b) = 
 \sum\limits_{\mbox{\tiny shuffles} \; \sigma} I(f_{\sigma(1)},f_{\sigma(2)},...,f_{\sigma(r)};a,b),
\end{eqnarray}
where the sum runs over all permutations $\sigma$, which preserve the relative order
of $1,2,...,k$ and of $k+1,...,r$.
\begin{figure}[t]
\sidecaption[t]
\includegraphics[bb= 150 650 420 725, scale=.65]{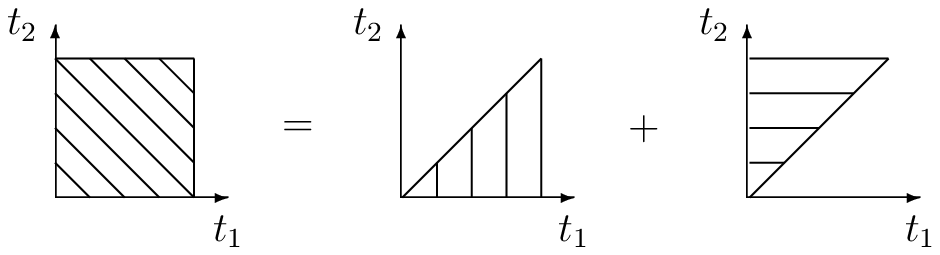}
\caption{A shuffle algebra follows from replacing the integral over the square by an integral over the lower triangle and an integral over the upper triangle.}
\label{weinzierl_fig6}
\end{figure}
The proof is sketched in fig.~\ref{weinzierl_fig6}.
The two outermost integrations are recursively replaced by integrations over the upper and lower triangle.
The definition of multiple polylogarithms in eq.~(\ref{weinzierl_Gfuncdef}) is of the form of iterated integrals as in eq.~(\ref{weinzierl_iterated_integrals}).
Therefore it follows that multiple polylogarithms obey a shuffle algebra.
An example for the multiplication is given by
\begin{eqnarray}
G(z_1;y) G(z_2;y) 
 & = & 
 G(z_1,z_2;y) + G(z_2,z_1;y).
\end{eqnarray}
Let us now turn to quasi-shuffle algebras.
Assume that for the set of letters we have an additional operation
\begin{eqnarray}
 (.,.) & : & A \otimes A \rightarrow A,
 \nonumber \\
       & &  l_1 \otimes l_2 \rightarrow (l_1, l_2),
\end{eqnarray}
which is commutative and associative.
Then we can define a new product of words recursively through
\begin{eqnarray}
\label{weinzierl_def_recursive_quasi_shuffle}
\left( l_1 l_2 ... l_k \right) \ast \left( l_{k+1} ... l_r \right) & = &
 l_1 \left[ \left( l_2 ... l_k \right) \ast \left( l_{k+1} ... l_r \right) \right]
+
 l_{k+1} \left[ \left( l_1 l_2 ... l_k \right) \ast \left( l_{k+2} ... l_r \right) \right]
 \nonumber \\
 & &
+
(l_1,l_{k+1}) \left[ \left( l_2 ... l_k \right) \ast \left( l_{k+2} ... l_r \right) \right].
\end{eqnarray}
This product is a generalisation of the shuffle product and differs from the recursive
definition of the shuffle product in eq.~(\ref{weinzierl_def_recursive_shuffle}) through the extra term in the last line.
This modified product is known under the names quasi-shuffle product \cite{Hoffman},
mixable shuffle product \cite{Guo}, stuffle product \cite{Borwein} or mould symmetrel \cite{Ecalle}.
Quasi-shuffle algebras are Hopf algebras.
Comultiplication and counit are defined as for the shuffle algebras.
The counit $\bar{e}$ is given by:
\begin{eqnarray}
\bar{e}\left( e\right) = 1, \;\;\;
& &
\bar{e}\left( l_1 l_2 ... l_n\right) = 0.
\end{eqnarray}
The coproduct $\Delta$ is given by:
\begin{eqnarray}
\Delta\left( l_1 l_2 ... l_k \right) 
& = & \sum\limits_{j=0}^k \left( l_{j+1} ... l_k \right) \otimes \left( l_1 ... l_j \right).
\end{eqnarray}
The antipode $S$ is recursively defined through
\begin{eqnarray}
S\left( l_1 l_2 ... l_k \right) & = & 
 - l_1 l_2 ... l_k
 - \sum\limits_{j=1}^{k-1} S\left( l_{j+1} ... l_k \right) \ast \left( l_1 ... l_j \right),
 \;\;\;\;\;\;
 S(e) = e.
\end{eqnarray}
An example for a quasi-shuffle algebra are nested sums.
Let $n_a$ and $n_b$ be integers with $n_a<n_b$ and let $f_1$, $f_2$, ... be functions
defined on the integers.
We consider the following nested sums:
\begin{eqnarray}
\label{weinzierl_nested_sums}
 S(f_1,f_2,...,f_k;n_a,n_b) 
 & = &
 \sum\limits_{i_1=n_a}^{n_b} f_1(i_1) \sum\limits_{i_2=n_a}^{i_1-1} f_2(i_2) 
 ...
 \sum\limits_{i_k=n_a}^{i_{k-1}-1} f_k(i_k).
\end{eqnarray}
(The letter $S$ denotes here a function, and not the antipode.)
For fixed $n_a$ and $n_b$ we have a quasi-shuffle algebra:
\begin{eqnarray}
\label{weinzierl_quasi_shuffle_multiplication}
\lefteqn{
 S(f_1,f_2,...,f_k;n_a,n_b) \ast S(f_{k+1},...,f_r; n_a,n_b) 
= } & &
 \nonumber \\
 & &
   \sum\limits_{i_1=n_a}^{n_b} f_1(i_1) \; S(f_2,...,f_k;n_a,i_1-1) \ast S(f_{k+1},...,f_r; n_a,i_1-1)
 \nonumber \\
 & &
 +  \sum\limits_{j_1=n_a}^{n_b} f_k(j_1) \; S(f_1,f_2,...,f_k;n_a,j_1-1) \ast S(f_{k+2},...,f_r; n_a,j_1-1)
 \nonumber \\
 & &
 +  \sum\limits_{i=n_a}^{n_b} f_1(i) f_k(i) \; S(f_2,...,f_k;n_a,i-1) \ast S(f_{k+2},...,f_r; n_a,i-1)
\end{eqnarray}
Note that the product of two letters corresponds to the point-wise product of the two functions:
\begin{eqnarray}
 ( f_i, f_j ) \; (n) & = & f_i(n) f_j(n).
\end{eqnarray}
The proof that nested sums obey the quasi-shuffle algebra is sketched in fig.~(\ref{weinzierl_fig5}).
The outermost sums of the nested sums on the l.h.s of (\ref{weinzierl_quasi_shuffle_multiplication}) are split into the three
regions indicated in fig.~(\ref{weinzierl_fig5}).
The definition of multiple polylogarithms in eq.~(\ref{weinzierl_def_multiple_polylogs_sum}) 
is of the form of nested sums as in eq.~(\ref{weinzierl_nested_sums}).
Therefore it follows that multiple polylogarithms obey also a quasi-shuffle algebra.
An example for the quasi-shuffle multiplication is given by
\begin{eqnarray}
 \mbox{Li}_{m_1}(x_1) \mbox{Li}_{m_2}(x_2) 
& = & 
 \mbox{Li}_{m_1,m_2}(x_1,x_2) + \mbox{Li}_{m_2,m_1}(x_2,x_1)
                                 + \mbox{Li}_{m_1+m_2}(x_1x_2).
\end{eqnarray}

\subsection{Mellin-Barnes transformation}
\label{weinzierl_mellin_barnes}

\index{Mellin-Barnes}
In sect.~\ref{weinzierl_section_feynman_integrals} we saw that the Feynman parameter integrals 
depend on two graph polynomials ${\mathcal U}$ and ${\mathcal F}$, which are homogeneous functions of the Feynman parameters.
In this section we will continue the discussion how these integrals can be performed and exchanged against a (multiple) sum over residues.
The case, where the two polynomials are absent is particular simple:
\begin{eqnarray}
\label{weinzierl_multi_beta_fct}
 \int\limits_{\Delta}  \omega
 \left( \prod\limits_{j=1}^n x_j^{\nu_j-1} \right)
 & = & 
 \frac{\prod\limits_{j=1}^{n}\Gamma(\nu_j)}{\Gamma(\nu_1+...+\nu_n)}.
\end{eqnarray}
With the help of the Mellin-Barnes transformation we now reduce the general case to eq.~(\ref{weinzierl_multi_beta_fct}).
The Mellin-Barnes transformation reads
\begin{eqnarray}
\label{weinzierl_multi_mellin_barnes}
\lefteqn{
\left(A_1 + A_2 + ... + A_n \right)^{-c} 
 = 
 \frac{1}{\Gamma(c)} \frac{1}{\left(2\pi i\right)^{n-1}} 
 \int\limits_{-i\infty}^{i\infty} d\sigma_1 ... \int\limits_{-i\infty}^{i\infty} d\sigma_{n-1}
 } & & \\
 & & 
 \times 
 \Gamma(-\sigma_1) ... \Gamma(-\sigma_{n-1}) \Gamma(\sigma_1+...+\sigma_{n-1}+c)
 \; 
 A_1^{\sigma_1} ...  A_{n-1}^{\sigma_{n-1}} A_n^{-\sigma_1-...-\sigma_{n-1}-c}.
 \nonumber 
\end{eqnarray}
Each contour is such that the poles of $\Gamma(-\sigma)$ are to the right and the poles
of $\Gamma(\sigma+c)$ are to the left.
This transformation can be used to convert the sum of monomials of the polynomials ${\mathcal U}$ and ${\mathcal F}$ into
a product, such that all Feynman parameter integrals are of the form of eq.~(\ref{weinzierl_multi_beta_fct}).
As this transformation converts sums into products it is 
the ``inverse'' of Feynman parametrisation.
With the help of eq.~(\ref{weinzierl_multi_beta_fct}) we may perform the integration over the Feynman parameters.
A single contour integral is then of the form
\begin{eqnarray}
\label{weinzierl_MellinBarnesInt}
I
 & = & 
\frac{1}{2\pi i} \int\limits_{-i\infty}^{+i\infty}
 d\sigma \; 
 \frac{\Gamma(\sigma+a_1) ... \Gamma(\sigma+a_m)}
      {\Gamma(\sigma+c_2) ... \Gamma(\sigma+c_p)}
 \frac{\Gamma(-\sigma+b_1) ... \Gamma(-\sigma+b_n)}
      {\Gamma(-\sigma+d_1) ... \Gamma(-\sigma+d_q)} 
 \; x^{-\sigma}.
\end{eqnarray}
The contour is such that the poles of $\Gamma(\sigma+a_1)$, ..., $\Gamma(\sigma+a_m)$ are to the right of the contour,
whereas the poles of $\Gamma(-\sigma+b_1)$,  ..., $\Gamma(-\sigma+b_n)$ are to the left of the contour.
We define
\begin{eqnarray}
& & \alpha = m+n-p-q,
\;\;\;\;\;\;
\beta = m-n-p+q, 
\nonumber \\
& & \lambda = \mbox{Re} \left( \sum\limits_{j=1}^m a_j
                              +\sum\limits_{j=1}^n b_j
                              -\sum\limits_{j=1}^p c_j
                              -\sum\limits_{j=1}^q d_j \right)
              - \frac{1}{2} \left( m+n-p-q \right).
\end{eqnarray}
Then the integral eq.~(\ref{weinzierl_MellinBarnesInt})
converges absolutely for $\alpha >0$ \cite{Erdelyi} and defines an analytic function in
\begin{eqnarray}
\left| \mbox{arg} \; x \right| & < & \mbox{min}\left( \pi, \alpha \frac{\pi}{2} \right).
\end{eqnarray}
The integral eq.~(\ref{weinzierl_MellinBarnesInt}) is most conveniently evaluated with 
the help of the residuum theorem by closing the contour to the left or to the right.
Therefore we need to know under which conditions the semi-circle at infinity used to close the contour gives a vanishing contribution.
This is obviously the case for $|x|<1$ if we close the contour to the left,
and for $|x|>1$, if we close the contour to the right.
The case $|x|=1$ deserves some special attention. One can show that
in the case $\beta=0$ the semi-circle gives a vanishing contribution, provided $\lambda < -1$.
To sum up all residues which lie inside the contour
it is useful to know the residues of the Gamma function:
\begin{eqnarray}
\mbox{res} \; \left( \Gamma(\sigma+a), \sigma=-a-n \right) = \frac{(-1)^n}{n!}, 
 & &
\mbox{res} \; \left( \Gamma(-\sigma+a), \sigma=a+n \right) = -\frac{(-1)^n}{n!}. 
 \nonumber
\end{eqnarray}
In the general case, the multiple integrals in eq.~(\ref{weinzierl_multi_mellin_barnes}) lead to multiple sums over residues. 

\subsection{Z-sums}
\label{weinzierl_Z_sums}

\index{nested sums}
The multiple sums over the residues can be expanded as a Laurent series in the dimensional regularisation parameter $\varepsilon$.
For particular integrals the coefficients of the Laurent series can be expressed in terms of multiple polylogarithms.
To see this, we first introduce a special form of nested sums, called 
$Z$-sums \cite{Moch:2001zr,Weinzierl:2002hv,Weinzierl:2004bn,Moch:2005uc}:
\begin{eqnarray} 
\label{weinzierl_definition_Zsum}
  Z(n;m_1,...,m_k;x_1,...,x_k) & = & \sum\limits_{i_1>i_2>\ldots>i_k>0}^n
     \frac{x_1^{i_1}}{{i_1}^{m_1}}\ldots \frac{x_k^{i_k}}{{i_k}^{m_k}}.
\end{eqnarray}
$k$ is called the depth of the $Z$-sum and $w=m_1+...+m_k$ is called the weight.
If the sums go to infinity ($n=\infty$) the $Z$-sums are multiple polylogarithms:
\begin{eqnarray}
\label{weinzierl_multipolylog}
Z(\infty;m_1,...,m_k;x_1,...,x_k) & = & \mbox{Li}_{m_1,...,m_k}(x_1,...,x_k).
\end{eqnarray}
For $x_1=...=x_k=1$ the definition reduces to the Euler-Zagier sums \cite{Euler,Zagier,Vermaseren:1998uu,Blumlein:1998if,Blumlein:2003gb}:
\begin{eqnarray}
Z(n;m_1,...,m_k;1,...,1) & = & Z_{m_1,...,m_k}(n).
\end{eqnarray}
For $n=\infty$ and $x_1=...=x_k=1$ the sum is a multiple $\zeta$-value:
\index{multiple zeta values}
\begin{eqnarray}
Z(\infty;m_1,...,m_k;1,...,1) & = & \zeta_{m_1,...,m_k}.
\end{eqnarray}
The $Z$-sums are of the form as in eq.~(\ref{weinzierl_nested_sums}) and form therefore a quasi-shuffle algebra.
The usefulness of the $Z$-sums lies in the fact, that they interpolate between multiple polylogarithms and Euler-Zagier sums.
Euler-Zagier sums appear in the expansion of the Gamma-function:
\index{series expansion}
\begin{eqnarray}
\label{weinzierl_expansiongamma}
\lefteqn{
\Gamma(n+\varepsilon)  = 
\Gamma(1+\varepsilon) \Gamma(n)
} & & \\
 & & 
 \times \left[
        1 + \varepsilon Z_1(n-1) + \varepsilon^2 Z_{11}(n-1)
          + \varepsilon^3 Z_{111}(n-1) + ... + \varepsilon^{n-1} Z_{11...1}(n-1)
 \right].
 \nonumber
\end{eqnarray}
The quasi-shuffle product can be used to reduce any product
\begin{eqnarray}
Z(n;m_1,...;x_1,...) \cdot Z(n;m_1',...;x_1',...)
\end{eqnarray}
of $Z$-sums with the same upper summation index $n$ to a linear combination of single $Z$-sums.
The Hopf algebra of $Z$-sums has additional structures if we allow expressions
\index{symbolic summation}
of the form \cite{Moch:2001zr}
\begin{eqnarray}
\label{weinzierl_augmented}
\frac{x_0^n}{n^{m_0}} Z(n;m_1,...,m_k;x_1,...,x_k),
\end{eqnarray}
e.g. $Z$-sums multiplied by a letter.
Then the following convolution product
\begin{eqnarray}
\label{weinzierl_convolution}
 \sum\limits_{i=1}^{n-1} \; \frac{x^i}{i^m} Z(i-1;...)
                         \; \frac{y^{n-i}}{(n-i)^{m'}} Z(n-i-1;...)
\end{eqnarray}
can again be expressed in terms of expressions of the form (\ref{weinzierl_augmented}).
In addition there is a conjugation, e.g. sums of the form 
\begin{eqnarray}
\label{weinzierl_conjugation}
 - \sum\limits_{i=1}^n 
       \left( \begin{array}{c} n \\ i \\ \end{array} \right)
       \left( -1 \right)^i
       \; \frac{x^i}{i^m} Z(i;...)
\end{eqnarray}
can also be reduced to terms of the form (\ref{weinzierl_augmented}).
The name conjugation stems from the following fact:
To any function $f(n)$ of an integer variable $n$ one can define
a conjugated function $C \circ f(n)$ as the following sum
\begin{eqnarray}
C \circ f(n) & = & \sum\limits_{i=1}^n 
       \left( \begin{array}{c} n \\ i \\ \end{array} \right)
       (-1)^i f(i).
\end{eqnarray}
Then conjugation satisfies the following two properties:
\begin{eqnarray}
C \circ 1 & = & 1,
 \nonumber \\
C \circ C \circ f(n) & = & f(n).
\end{eqnarray}
Finally there is the combination of conjugation and convolution,
e.g. sums of the form 
\begin{eqnarray}
\label{weinzierl_conjugationconvolution}
 - \sum\limits_{i=1}^{n-1} 
       \left( \begin{array}{c} n \\ i \\ \end{array} \right)
       \left( -1 \right)^i
       \; \frac{x^i}{i^m} Z(i;...)
       \; \frac{y^{n-i}}{(n-i)^{m'}} Z(n-i;...)
\end{eqnarray}
can also be reduced to terms of the form (\ref{weinzierl_augmented}).

With the help of these algorithms it is possible to prove that the Laurent series in $\varepsilon$ of specific Feynman integrals contains only
multiple polylogarithms \cite{Bierenbaum:2003ud}.

% -----------------------------------------------------------------------------------
\section{Beyond multiple polylogarithms}
\label{weinzierl_section_beyond_multiple_polylogs}

\index{differential equation}
\index{Picard-Fuchs equation}
Although multiple polylogarithms form an important class of functions, which appear in the evaluation
of Feynman integrals, it is known from explicit calculations that starting from two-loop integrals with massive particles 
one encounters functions beyond the  class of multiple polylogarithms.
The simplest example is given by the two-loop sunset diagram with non-zero masses, where elliptic integrals make their appearance \cite{Laporta:2004rb}.
Differential equations provide a tool to get a handle on these integrals \cite{Kotikov:1990kg,Kotikov:1991pm,Remiddi:1997ny,Gehrmann:1999as,Gehrmann:2000zt,Gehrmann:2001ck,Argeri:2007up,MullerStach:2012mp}.

\subsection{The two-loop sunset integral with non-zero masses}
\label{weinzierl_sunset}

In this subsection we review how ideas of algebraic geometry can be used to obtain a differential equation for the Feynman integral \cite{MullerStach:2011ru}.
We first focus on the example of the two-loop sunrise integral.
The two-loop sunrise integral is given in $D$-dimensional Minkowski space by
\begin{eqnarray}
\label{weinzierl_def_sunrise}
 S\left( D, p^2 \right)
 = 
 \frac{\left(\mu^2\right)^{3-D}}{\Gamma\left(3-D\right)}
 \int \frac{d^Dk_1}{i \pi^{\frac{D}{2}}} \frac{d^Dk_2}{i \pi^{\frac{D}{2}}}
 \frac{1}{\left(-k_1^2+m_1^2\right)\left(-k_2^2+m_2^2\right)\left(-k_3^2+m_3^2\right)},
\end{eqnarray}
with $k_3=p-k_1-k_2$.
Here we suppressed on the l.h.s. the dependence on the internal masses $m_1$, $m_2$ and $m_3$ and on the arbitrary scale $\mu$.
It is convenient to denote the momentum squared by $t=p^2$.
In terms of Feynman parameters the integral reads
\begin{eqnarray}
\label{weinzierl_def_Feynman_sunrise_integral}
 S\left( D, t\right)
 & = & 
 \int\limits_{\Delta} \omega \frac{{\cal U}^{3-\frac{3}{2}D}}{{\cal F}^{3-D}},
\end{eqnarray}
where the two Feynman graph polynomials are given by
\begin{eqnarray}
 {\cal F} = \left[ - x_1 x_2 x_3 t
                + \left( x_1 m_1^2 + x_2 m_2^2 + x_3 m_3^2 \right) {\cal U} \right] \mu^{-2},
 & &
 {\cal U} = x_1 x_2 + x_2 x_3 + x_3 x_1.
 \nonumber 
\end{eqnarray}
It is simpler to consider this integral first in $D=2$ dimensions and to obtain the result in $D=4-2\varepsilon$ dimensions with the help 
of dimensional recurrence relations \cite{Tarasov:1996br,Tarasov:1997kx}.
In two dimensions this integral is finite, depends only on the second Symanzik polynomial ${\cal F}$ 
and is given by
\begin{eqnarray}
\label{weinzierl_def_Feynman_integral_dim_two}
 S\left( 2, t\right)
 & = & 
 \int\limits_{\Delta} \frac{\omega}{{\cal F}}.
\end{eqnarray}
From the point of view of algebraic geometry there are two objects of interest in eq.~(\ref{weinzierl_def_Feynman_integral_dim_two}):
On the one hand the domain of integration $\Delta$ and on the other hand the algebraic variety $X$ defined by the zero set of ${\cal F}=0$.
The two objects $X$ and $\Delta$ intersect at the three points
$[1:0:0]$, $[0:1:0]$ and $[0:0:1]$ of the projective space ${\mathbb P}^2$.
\begin{figure}[t]
\sidecaption[t]
\includegraphics[bb= 230 585 365 740, scale=.65]{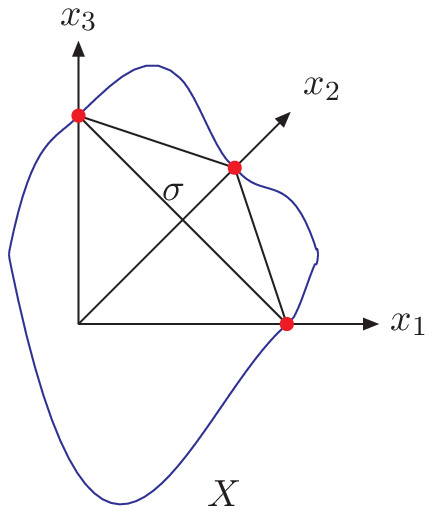}
\includegraphics[bb= 190 570 345 730, scale=.65]{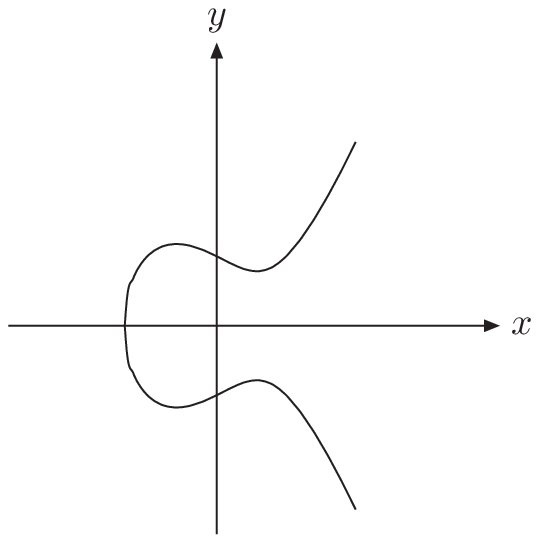}
\caption{The intersection of the domain of integration $\Delta$ with the zero set $X$ of the second Symanzik polynomial (left) and
the elliptic curve $y^2=x^3-x+1$ (right).}
\label{weinzierl_fig7}
\end{figure}
This is shown in fig.~(\ref{weinzierl_fig7}) on the left.
We blow-up ${\mathbb P}^2$ in these three points and we denote the blow-up by $P$.
We further denote the strict transform of $X$ by $Y$ and the total transform of the set 
$\{ x_1 x_2 x_3 = 0 \}$ by $B$.
With these notations we can now consider the mixed Hodge structure (or the motive) given by the relative cohomology group \cite{Bloch:2006}
\begin{eqnarray}
 H^2\left(P \backslash Y, B \backslash B \cap Y \right).
\end{eqnarray}
In the case of the two-loop sunrise integral considered here essential information on 
$H^2(P \backslash Y, B \backslash B \cap Y )$ is already given by $H^1(X)$.
We recall that the algebraic variety $X$ is defined by the second Symanzik polynomial:
\begin{eqnarray}
 - x_1 x_2 x_3 t + \left( x_1 m_1^2 + x_2 m_2^2 + x_3 m_3^2 \right) \left( x_1 x_2 + x_2 x_3 + x_3 x_1 \right) & = & 0.
\end{eqnarray}
This defines for generic values of the parameters $t$, $m_1$, $m_2$ and $m_3$ an elliptic curve.
The elliptic curve varies smoothly with the parameters $t$, $m_1$, $m_2$ and $m_3$.
By a birational change of coordinates this equation can brought into the Weierstrass normal form
\begin{eqnarray}
 y^2 z - x^3 - a_2(t) x z^2 - a_3(t) z^3 & = & 0.
\end{eqnarray}
The dependence of $a_2$ and $a_3$ on the masses is not written explicitly.
In the chart $z=1$ this reduces to
\begin{eqnarray}
\label{weierstrass_normal_form}
 y^2 - x^3 - a_2(t) x - a_3(t) & = & 0.
\end{eqnarray}
The curve varies with the parameter $t$.
An example of an elliptic curve is shown in fig.~(\ref{weinzierl_fig7}) on the right.
It is well-known that in the coordinates of eq.~(\ref{weierstrass_normal_form}) the cohomology group $H^1(X)$ 
is generated by
\begin{eqnarray}
 \eta = \frac{dx}{y}
 & \mbox{and} &
 \dot{\eta} = \frac{d}{dt} \eta.
\end{eqnarray}
Since $H^1(X)$ is two-dimensional it follows that $\ddot{\eta}=\frac{d^2}{dt^2} \eta$ must be a linear combination of $\eta$ and $\dot{\eta}$.
In other words we must have a relation of the form
\begin{eqnarray}
 p_0(t) \ddot{\eta} + p_1(t) \dot{\eta} + p_2(t) \eta & = & 0.
\end{eqnarray}
The coefficients $p_0(t)$, $p_1(t)$ and $p_2(t)$ define the Picard-Fuchs operator
\begin{eqnarray}
 L^{(2)} & = & p_0(t) \frac{d^2}{dt^2} + p_1(t) \frac{d}{dt} + p_2(t).
\end{eqnarray}
Applying the Picard-Fuchs operator to our integrand gives an exact form:
\begin{eqnarray}
 L^{(2)} \left(  \frac{\omega}{{\cal F}} \right)
 & = & 
 d \beta.
\end{eqnarray}
The integration over $\Delta$ yields
\begin{eqnarray}
 L^{(2)} S(2,t) & = & \int\limits_\Delta d \beta = \int\limits_{\partial \Delta} \beta
\end{eqnarray}
The integration of $\beta$ over $\partial \Delta$ is elementary and we arrive at
\begin{eqnarray}
\label{final_result}
 \left[ p_0(t) \frac{d^2}{d t^2} + p_1(t) \frac{d}{dt} + p_2(t)  \right] S\left(2,t\right) & = & p_3(t).
\end{eqnarray}
This is the sought-after second-order differential equation. The coefficients are given in the equal mass case by \cite{Laporta:2004rb,Broadhurst:1993mw}
\begin{align}
 & 
 p_0(t) = 
  t \left( t - m^2 \right)
    \left( t - 9 m^2 \right),
 & &
 p_2(t) = 
  t - 3 m^2,
 \nonumber \\
 &
 p_1(t) = 
  3 t^2 - 20 t m^2 + 9 m^4, 
 & &
 p_3(t) = 
 - 6 \mu^2.
\end{align}
The coefficients for the unequal mass case can be found in \cite{MullerStach:2011ru}.

\subsection{Differential equations}
\label{weinzierl_differential_equations}

The ideas of the previous subsection can be generalised to arbitrary Feynman integrals.
For a given Feynman integral let us pick one variable $t$ from the set of the Lorentz invariant quantities $( p_j + p_k )^2$ and the 
internal masses squared $m_i^2$.
Let us write
\begin{eqnarray}
 \omega_t & = & 
 \omega
 \left( \prod\limits_{j=1}^n x_j^{\nu_j-1} \right)
 \frac{{\mathcal U}^{\nu-(l+1) D/2}}{{\mathcal F}^{\nu-l D/2}}.
\end{eqnarray}
The subscript $t$ indicates that $\omega_t$ depends on $t$ through ${\mathcal F}$. 
The Feynman integral is then simply
\begin{eqnarray}
I_G  & = &
 \int\limits_{\Delta}  \omega_t
\end{eqnarray}
We seek an ordinary linear differential equation with respect to the variable $t$ for the Feynman integral $I_G$.
We start to look for a differential equation of the form
\begin{eqnarray}
\label{weinzierl_basic_picard}
 L^{(r)} \omega_t & = & d \beta,
\end{eqnarray}
where
\begin{eqnarray}
\label{weinzierl_picard_fuchs_operator}
 L^{(r)} & = & \sum\limits_{j=0}^r p_j \left( \mu^2 \frac{d}{dt} \right)^j
\end{eqnarray}
is a Picard-Fuchs operator of order $r$.
Suppose an equation of the form as in eq.~(\ref{weinzierl_basic_picard}) exists.
Following the same steps as in section~(\ref{weinzierl_sunset}) we arrive at
\begin{eqnarray}
 L^{(r)} I_G & = & \int\limits_{\partial \Delta} \beta.
\end{eqnarray}
The right-hand side corresponds to simpler Feynman integrals, where one propagator has been contracted.
The coefficients of the Picard-Fuchs operator and the coefficients of the form $\beta$ can be found by solving
a linear system of equations \cite{MullerStach:2012mp}.

% -----------------------------------------------------------------------------------
%\bibliographystyle{spmpsci}
%\bibliography{/home/stefanw/notes/biblio}

\begin{thebibliography}{10}
\providecommand{\url}[1]{{#1}}
\providecommand{\urlprefix}{URL }
\expandafter\ifx\csname urlstyle\endcsname\relax
  \providecommand{\doi}[1]{DOI~\discretionary{}{}{}#1}\else
  \providecommand{\doi}{DOI~\discretionary{}{}{}\begingroup
  \urlstyle{rm}\Url}\fi

\bibitem{Ablinger:2011te}
Ablinger, J., Blumlein, J., Schneider, C.: {Harmonic Sums and Polylogarithms
  Generated by Cyclotomic Polynomials}.
\newblock J.Math.Phys. \textbf{52}, 102,301 (2011)

\bibitem{Argeri:2007up}
Argeri, M., Mastrolia, P.: {Feynman Diagrams and Differential Equations}.
\newblock Int. J. Mod. Phys. \textbf{A22}, 4375--4436 (2007)

\bibitem{Bierenbaum:2003ud}
Bierenbaum, I., Weinzierl, S.: The massless two-loop two-point function.
\newblock Eur. Phys. J. \textbf{C32}, 67--78 (2003)

\bibitem{Bloch:2006}
Bloch, S., Esnault, H., Kreimer, D.: On motives associated to graph
  polynomials.
\newblock Commun. Math. Phys. \textbf{267}, 181 (2006)

\bibitem{Blumlein:2003gb}
Bl{\"u}mlein, J.: Algebraic relations between harmonic sums and associated
  quantities.
\newblock Comput. Phys. Commun. \textbf{159}, 19--54 (2004)

\bibitem{Blumlein:2009}
Bl{\"u}mlein, J., Broadhurst, D.J., Vermaseren, J.A.M.: {The Multiple Zeta
  Value Data Mine}.
\newblock Comput. Phys. Commun. \textbf{181}, 582 (2010)

\bibitem{Blumlein:1998if}
Bl{\"u}mlein, J., Kurth, S.: Harmonic sums and mellin transforms up to two-loop
  order.
\newblock Phys. Rev. \textbf{D60}, 014,018 (1999)

\bibitem{Bogner:2010kv}
Bogner, C., Weinzierl, S.: {Feynman graph polynomials}.
\newblock Int. J. Mod. Phys. \textbf{A25}, 2585--2618 (2010)

\bibitem{Borwein}
Borwein, J.M., Bradley, D.M., Broadhurst, D.J., Lisonek, P.: Special values of
  multiple polylogarithms.
\newblock Trans. Amer. Math. Soc. \textbf{353:3}, 907 (2001)

\bibitem{Broadhurst:1993mw}
Broadhurst, D.J., Fleischer, J., Tarasov, O.: {Two loop two point functions
  with masses: Asymptotic expansions and Taylor series, in any dimension}.
\newblock Z.Phys. \textbf{C60}, 287--302 (1993)

\bibitem{Brown:2008}
Brown, F.: The massless higher-loop two-point function.
\newblock Commun. Math. Phys. \textbf{287}, 925--958 (2008)

\bibitem{Brown:2009a}
Brown, F.: {On the periods of some Feynman integrals}.
\newblock arXiv:0910.0114 [math.AG] (2009)

\bibitem{Chaiken:1982}
Chaiken, S.: A combinatorial proof of the all minors matrix tree theorem.
\newblock SIAM J. Alg. Disc. Meth. \textbf{3}, 319--329 (1982)

\bibitem{Chen:1982}
Chen, W.K.: Applied graph theory, graphs and electrical networks.
\newblock North Holland (1982)

\bibitem{Dodgson:1866}
Dodgson, C.L.: Condensation of determinants.
\newblock Proc. Roy. Soc. London \textbf{15}, 150--155 (1866)

\bibitem{Ecalle}
Ecalle, J.: Ari/gari, la dimorphie et l'arithm{\'e}tique des multiz{\^e}tas: un
  premier bilan.
\newblock Journal de Th{\'e}orie des Nombres de Bordeaux \textbf{15}, 411
  (2003)

\bibitem{Erdelyi}
Erd{\'e}lyi, A., Magnus, W., Oberhettinger, F., Tricomi, F.: Higher
  Transcendental Functions.
\newblock Vol. I, McGraw Hill (1953)

\bibitem{Euler}
Euler, L.: Meditationes circa singulare serierum genus.
\newblock Novi Comm. Acad. Sci. Petropol. \textbf{20}, 140 (1775)

\bibitem{Gehrmann:1999as}
Gehrmann, T., Remiddi, E.: Differential equations for two-loop four-point
  functions.
\newblock Nucl. Phys. \textbf{B580}, 485--518 (2000)

\bibitem{Gehrmann:2001ck}
Gehrmann, T., Remiddi, E.: Two-loop master integrals for gamma* {$\rightarrow$} 3jets: The
  non- planar topologies.
\newblock Nucl. Phys. \textbf{B601}, 287--317 (2001)

\bibitem{Gehrmann:2000zt}
Gehrmann, T., Remiddi, E.: Two-loop master integrals for gamma* {$\rightarrow$} 3jets: The
  planar topologies.
\newblock Nucl. Phys. \textbf{B601}, 248--286 (2001)

\bibitem{Goncharov_no_note}
Goncharov, A.B.: Multiple polylogarithms, cyclotomy and modular complexes.
\newblock Math. Res. Lett. \textbf{5}, 497 (1998)

\bibitem{Goncharov:2001}
Goncharov, A.B.: Multiple polylogarithms and mixed Tate motives.
\newblock math.AG/0103059 (2001)

\bibitem{Guo}
Guo, L., Keigher, W.: Baxter algebras and shuffle products.
\newblock Adv. in Math. \textbf{150}, 117 (2000)

\bibitem{Hoffman}
Hoffman, M.E.: Quasi-shuffle products.
\newblock J. Algebraic Combin. \textbf{11}, 49 (2000)

\bibitem{Kotikov:1991pm}
Kotikov, A.V.: Differential equation method: The calculation of n point Feynman
  diagrams.
\newblock Phys. Lett. \textbf{B267}, 123--127 (1991)

\bibitem{Kotikov:1990kg}
Kotikov, A.V.: Differential equations method: New technique for massive Feynman
  diagrams calculation.
\newblock Phys. Lett. \textbf{B254}, 158--164 (1991)

\bibitem{Laporta:2004rb}
Laporta, S., Remiddi, E.: {Analytic treatment of the two loop equal mass
  sunrise graph}.
\newblock Nucl. Phys. \textbf{B704}, 349--386 (2005)

\bibitem{Moch:2005uc}
Moch, S., Uwer, P.: Xsummer: Transcendental functions and symbolic summation in
  form.
\newblock Comput. Phys. Commun. \textbf{174}, 759--770 (2006)

\bibitem{Moch:2001zr}
Moch, S., Uwer, P., Weinzierl, S.: Nested sums, expansion of transcendental
  functions and multi-scale multi-loop integrals.
\newblock J. Math. Phys. \textbf{43}, 3363--3386 (2002)

\bibitem{Moon:1994}
Moon, J.: Some determinant expansions and the matrix-tree theorem.
\newblock Discrete Math. \textbf{124}, 163--171 (1994)

\bibitem{MullerStach:2011ru}
M{\"u}ller-Stach, S., Weinzierl, S., Zayadeh, R.: {A second-order differential
  equation for the two-loop sunrise graph with arbitrary masses}.
\newblock Commun. Num. Theor. Phys. \textbf{6}, 203--222 (2012)

\bibitem{MullerStach:2012mp}
M{\"u}ller-Stach, S., Weinzierl, S., Zayadeh, R.: {Picard-Fuchs equations for
  Feynman integrals}.
\newblock arXiv:1212.4389 [hep-ph] (2012)

\bibitem{Nielsen}
Nielsen, N.: Der Eulersche Dilogarithmus und seine Verallgemeinerungen.
\newblock Nova Acta Leopoldina (Halle) \textbf{90}, 123 (1909)

\bibitem{Remiddi:1997ny}
Remiddi, E.: Differential equations for Feynman graph amplitudes.
\newblock Nuovo Cim. \textbf{A110}, 1435--1452 (1997)

\bibitem{Remiddi:1999ew}
Remiddi, E., Vermaseren, J.A.M.: Harmonic polylogarithms.
\newblock Int. J. Mod. Phys. \textbf{A15}, 725 (2000)

\bibitem{Reutenauer}
Reutenauer, C.: Free Lie Algebras.
\newblock Clarendon Press, Oxford (1993)

\bibitem{Stanley:1998}
Stanley, R.P.: Spanning trees and a conjecture of Kontsevich.
\newblock Ann. Combin. \textbf{2}, 351--363 (1998)

\bibitem{Sweedler}
Sweedler, M.: Hopf Algebras.
\newblock Benjamin, New York (1969)

\bibitem{Tarasov:1996br}
Tarasov, O.V.: Connection between Feynman integrals having different values of
  the space-time dimension.
\newblock Phys. Rev. \textbf{D54}, 6479--6490 (1996)

\bibitem{Tarasov:1997kx}
Tarasov, O.V.: Generalized recurrence relations for two-loop propagator
  integrals with arbitrary masses.
\newblock Nucl. Phys. \textbf{B502}, 455--482 (1997)

\bibitem{Tutte:1984}
Tutte, W.T.: Graph Theory, \emph{Encyclopedia of mathematics and its
  applications}, vol.~21.
\newblock Addison-Wesley (1984)

\bibitem{Vermaseren:1998uu}
Vermaseren, J.A.M.: Harmonic sums, mellin transforms and integrals.
\newblock Int. J. Mod. Phys. \textbf{A14}, 2037 (1999)

\bibitem{Vollinga:2004sn}
Vollinga, J., Weinzierl, S.: Numerical evaluation of multiple polylogarithms.
\newblock Comput. Phys. Commun. \textbf{167}, 177 (2005)

\bibitem{Weinzierl:2002hv}
Weinzierl, S.: Symbolic expansion of transcendental functions.
\newblock Comput. Phys. Commun. \textbf{145}, 357--370 (2002)

\bibitem{Weinzierl:2004bn}
Weinzierl, S.: Expansion around half-integer values, binomial sums and inverse
  binomial sums.
\newblock J. Math. Phys. \textbf{45}, 2656--2673 (2004)

\bibitem{Zagier}
Zagier, D.: Values of zeta functions and their applications.
\newblock First European Congress of Mathematics, Vol. II, Birkhauser, Boston
  p. 497 (1994)

\bibitem{Zeilberger:1997}
Zeilberger, D.: Dodgson's determinant-evaluation rule proved by two-timing men
  and women.
\newblock Electron. J. Combin. \textbf{4}(R22), 2 (1997)

\end{thebibliography}

% -----------------------------------------------------------------------------------

\end{document}